\pdfoutput=1
\documentclass{sig-alt-gov1}
\usepackage{listings}
\usepackage{courier}
\begin{document}
\conferenceinfo{JCDL'07,} {June 17--22, 2007, Vancouver, British Columbia, Canada.}
\CopyrightYear{2007}
\crdata{978-1-59593-644-8/07/0006} 

\title{A Practical Ontology for the Large-Scale Modeling of Scholarly Artifacts and their Usage}

\numberofauthors{3} 
\author{
\alignauthor
Marko A. Rodriguez\\
       \affaddr{Digital Library Research \& Prototyping Team}\\
       \affaddr{Los Alamos National Laboratory}\\
       \affaddr{Los Alamos, NM 87545}\\
       \email{marko@lanl.gov}
\alignauthor
Johan Bollen\\
       \affaddr{Digital Library Research \& Prototyping Team}\\
       \affaddr{Los Alamos National Laboratory}\\
       \affaddr{Los Alamos, NM 87545}\\
       \email{jbollen@lanl.gov}
\alignauthor 
Herbert Van de Sompel\\
       \affaddr{Digital Library Research \& Prototyping Team}\\
       \affaddr{Los Alamos National Laboratory}\\
       \affaddr{Los Alamos, NM 87545}\\
       \email{herbertv@lanl.gov}}

\maketitle
\begin{abstract}
The large-scale analysis of scholarly artifact usage is constrained primarily by current practices in usage data archiving, privacy issues concerned with the dissemination of usage data, and the lack of a practical ontology for modeling the usage domain. As a remedy to the third constraint, this article presents a scholarly ontology that was engineered to represent those classes for which large-scale bibliographic and usage data exists, supports usage research, and whose instantiation is scalable to the order of 50 million articles along with their associated artifacts (e.g.~authors and journals) and an accompanying 1 billion usage events. The real world instantiation of the presented abstract ontology is a semantic network model of the scholarly community which lends the scholarly process to statistical analysis and computational support. We present the ontology, discuss its instantiation, and provide some example inference rules for calculating various scholarly artifact metrics.
\end{abstract}



\section{Introduction}

New publications are added to the scholarly record at an accelerating pace. This point is realized by observing the evolution of the amount of publications indexed in Thomson Scientific's citation database over the last fifteen years: 875,310 in 1990; 1,067,292 in 1995; 1,164,015 in 2000, and 1,511,067 in 2005.  However, the extent of the scholarly record reaches far beyond what is indexed by Thompson Scientific. While Thompson Scientific focuses primarily on quality-driven journals (roughly 8,700 in 2005), they do not index more novel scholarly artifacts such as preprints deposited in institutional or discipline-oriented repositories, datasets, software, and simulations that are increasingly being considered scholarly communication units in their own right.

While the size (and growth) of the scholarly record is impressive, the extent of its use is even more staggering.  For instance, in November 2006, Elsevier's Science Direct, which provides access to articles from approximately 2,000 journals, celebrated its 1 billionth full-text download since counting started in April of 1999\footnote{Elsevier's 1 billion downloads article available at: http://www.info.sciencedirect.com/news/archive/2006/ \newline news\_billionth.asp}. And, again, the extent of scholarly usage clearly reaches far beyond Elsevier's repository. Furthermore, usage events include not only full-text downloads, but also events such as requesting services from linking servers, downloading bibliographic citations, emailing abstracts, etc. 

To a large extent, the effect of usage behavior on the scholarly process is a horizon that is only beginning to be understood and, if properly studied, will offer clues to the evolutionary trends of science \cite{bibliom:kurtz2004,earlie:brody2006,mappin:bollen2006}, quantitative models of the value of scholarly artifacts \cite{altern:bollen2005,usage:bollen2006}, and services to support scholars \cite{bx:bollen2006}. The Andrew W. Mellon funded MESUR\footnote{MEtrics from Scholarly Usage of Resources available at: http://www.mesur.org/} project at the Research Library of the Los Alamos National Laboratory aims at developing metrics for assessing scholarly communication artifacts (e.g.~articles, journals, conference proceedings, etc.) and agents (e.g.~authors, institutions, publishers, repositories, etc.) on the basis of scholarly usage.  In order to do this, the MESUR project makes use of a representative collection of bibliographic, citation and usage data. This data is collected from a wide variety of sources including academic publishers, secondary publishers, institutional linking servers, etc.  Expectations are that the collected data will eventually encompass tens of millions of bibliographic records, hundreds of millions of citations, and billions of usage events.  Mining such a vast data set in an efficient, performing, and flexible manner presents significant challenges regarding data representation and data access. 
This article presents, the OWL ontology \cite{owlspec:mcguinness2004} used by MESUR to represent bibliographic, citation and usage data in an integrated manner. The proposed MESUR ontology is practical, as opposed to all encompassing, in that it represents those artifacts and properties that, as previously shown in \cite{bx:bollen2006}, are realistically available from modern scholarly information systems. This includes bibliographic data such as author, title, identifier, publication date and usage data such as the IP address of the accessing agent, the date and time of access, type of usage, etc.  Finally, another novel contribution of this work is the hybrid storage and access architecture in which relational database and triple store technology are combined. This is achieved by storing core data and relationships in the triple store and auxiliary data in a relational database. This design choice is driven by the need to keep the size of the triple store to a level that can realistically be handled by current technologies. The combination of the data architecture and scholarly ontology presented in this article provide the foundation for the large-scale modeling and analysis of scholarly artifacts and their usage.

\section{Semantic Network Ontologies}

A semantic network (sometimes called a multi-relational network or multi-graph) is composed of a set of nodes (representing heterogeneous artifacts) connected to one another by a set of qualified, or labeled, edges \cite{sowa:semantic1991}. In a graph theoretic sense, a semantic network is a directed labeled graph. Because an edge is labeled, two nodes can be connected to one another by an infinite number of edges. However, in most cases, the possible interconnections between node types is constrained to a predetermined set. This predetermined set is made explicit in the semantic network's associated ontology. An ontology is generally defined as a set of abstract classes, their relationship to one another, and a collection of inference rules for deriving implicit relationships \cite{devsem:alesso2005}. An ontology makes no explicit reference to the actual instances of the defined abstract classes; this is the role of the semantic network.

An ontology is related to the developer's API in object oriented programming languages such as C++ and Java (minus the explicit representation of class methods/functions). For example, the set of relationships of an ontological class are known as the class' properties and, in the object oriented lexicon, can be understood as class fields. Also, a taxonomy is usually expressed in a semantic network ontology. A taxonomy of sub- and super-classes support the inheritance of class properties. For instance, if all mammals are warm blooded, then all humans are warm blooded because all humans are mammals. In an inheritance hierarchy, the warm blooded property of mammals is inherited by all sub-classes of mammal (e.g.~human).

\begin{figure}[h!]
	\centering
	\includegraphics[width=0.3\textwidth]{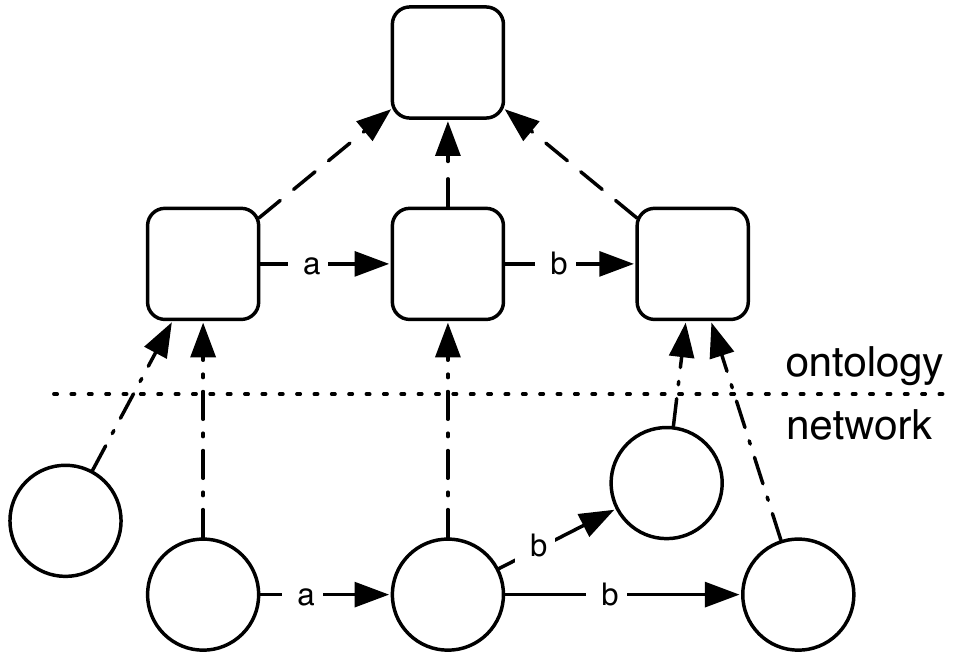}
	 \caption{The relationship between an ontology and its semantic network instantiation \label{fig:ont-inst-demo}}
\end{figure}

Figure \ref{fig:ont-inst-demo} diagrams the relationship between an ontology and its semantic network instantiation. The circles represents objects that are instances of the dash-dot pointed to abstract classes (the squares). The three lower squares are subclasses of a more general top-level class (denoted by the dashed edges). The horizontal edges in the ontology denote permissible property types in the instantiation and thus, corresponding horizontal labeled edges in the semantic network may exist. Figure \ref{fig:ont-inst-demo} does not expose the range of conceptual nuances that can be expressed by modern ontology languages and thus, only provides a rudimentary representation of the relationship between an ontology and its semantic network instantiation.

\subsection{Semantic Network Technology}

The most popular semantic network representational framework is the Resource Description Framework and Schema, or RDF(S) \cite{rdfspec:manola2004}. RDF(S) represents all nodes and edges by Universal Resource Identifiers (URI) \cite{uri:berners2005}. The URI approach supports the use of namespacing such that the URI \texttt{http://www.science.org\#Article} has a different meaning, or connotation, than what may be understood by the URI \texttt{http://www.newspaper.net\#Article}.

The Web Ontology Language (OWL) is an extension of RDF(S) that supports a richer vocabulary (e.g.~promotes many set theoretical concepts) \cite{owlspec:mcguinness2004}. Prot\'{e}g\'{e}\footnote{Prot\'{e}g\'{e} available at: http://protege.stanford.edu/} is perhaps the most popular application for designing OWL ontologies \cite{protege:noy2000}. While OWL is primarily a machine readable language, an OWL ontology can be diagrammed using the Unified Modeling Language's (UML) class diagrams (i.e.~entity relationship diagrams).

Modern semantic network data stores represent the relationship between two nodes by a \textit{triple}. For instance, the triple
\begin{equation*}
 \langle \text{URI}_a, \; \texttt{http://xmlns.com/foaf/0.1/\#knows}, \; \text{URI}_b \rangle
 \end{equation*}
 \begin{figure}[ht!]
	\centering
	\includegraphics[width=0.4\textwidth]{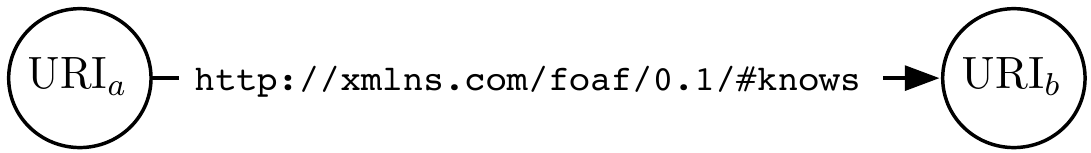}
	 \caption{A diagrammed triple \label{fig:foaf-mini}}
\end{figure}
 states that the resource identified by $\text{URI}_a$ knows the resource identified by $\text{URI}_b$, where $\text{URI}_a$ and $\text{URI}_b$ are nodes and \texttt{http://xmlns.com/foaf/0.1/\#knows} is a directed labeled edge (see Figure \ref{fig:foaf-mini}). The meaning of \texttt{knows} is fully defined by the URI \texttt{http://xmlns.com/foaf/0.1/}. The union of instantiated FOAF triples is a FOAF semantic network. Current platforms for storing and querying such semantic networks are called \textit{triple stores}. Many open source and proprietary triple stores currently exist. Various querying languages exist as well \cite{semtech:Mag2002}. The role of the query language is to provide the interface to access the data contained in the triple store. This is analogous to the relationships between SQL and relational databases. Perhaps the most popular triple store query language is SPARQL \cite{sparql:prud2004}. An example SPARQL query is
 
\begin{footnotesize}
\begin{verbatim}
SELECT  ?x
WHERE   ( ?x foaf:knows vub:cgershen ).
\end{verbatim}
\end{footnotesize}

In the above query, the \texttt{?x} variable is bound to any node that is the domain of a triple with an associated predicate of \texttt{http://xmlns.com/foaf/0.1/\#knows} and a range of \texttt{http://homepages.vub.ac.be/\#cgershen}. Thus, the above query returns all people who know \texttt{vub:cgershen} (i.e.~Carlos Gershenson). 
 
\begin{figure}[h!]
	\centering
	\includegraphics[width=0.375\textwidth]{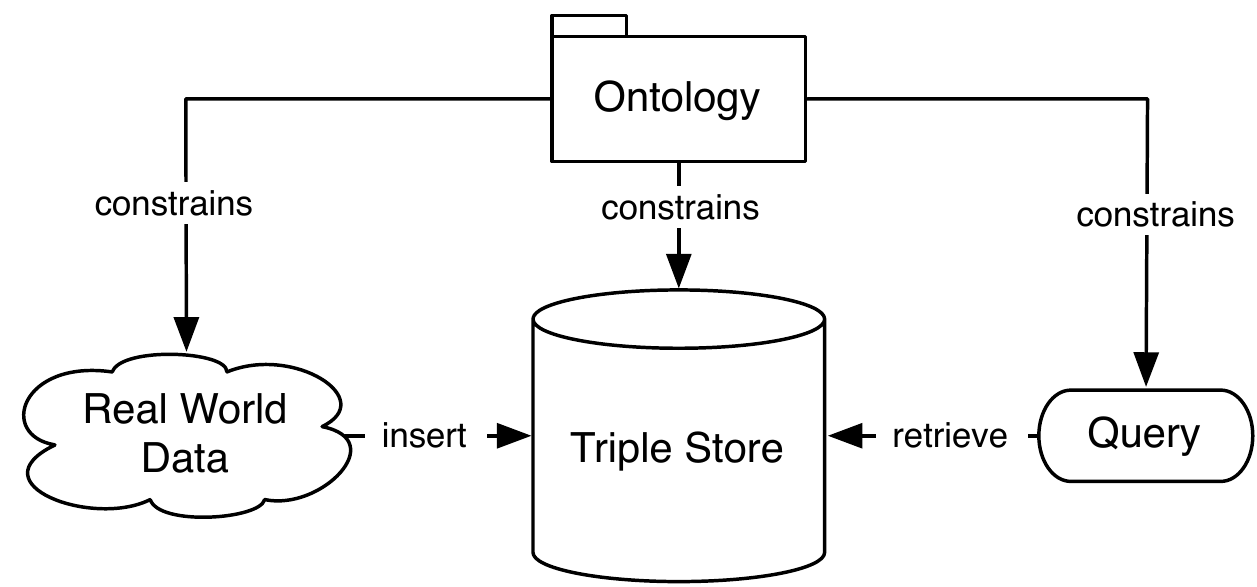}
	 \caption{The many roles of an ontology \label{fig:full-system}}
\end{figure}

The ontology plays a significant role in many aspects of a semantic network. Figure \ref{fig:full-system} demonstrates the role of the ontology in determining which real world data is harvested, how that data is represented inside of the triple store (semantic network), and finally, what queries and inferences are possible to execute.

\section{Scholarly Ontologies}

In general, an ontology's classes, their relationships, and inferences are determined according to what is being modeled, for what problems that model is trying to solve, and how that model's classes can be instantiated according to real world data. Thus, there were three primary requirements to the development of the MESUR ontology:

\begin{enumerate}
\item realistically available real world data
\item ability to study usage behavior
\item scalability of the triple store instantiation.
\end{enumerate}

Without real-world data, an ontology serves only as a conceptual tool for understanding a particular domain and, in such cases, ontologies of this nature may be very detailed in what they represent. However, for ontologies that are designed to be instantiated by real world data, the ontology is ultimately constrained by data availability. Thus, the MESUR ontology is constrained to bibliographic and usage data since these are the primary sources of scholarly data. In the scholarly community, while articles, journals, conference proceedings, and the like are well documented and represented in formats that lend themselves to analysis, other information, such as usage data, tends to be less explicit due to the inherent privacy issues surrounding individual usage behavior. Therefore, a primary objective of the MESUR project is the acquisition of large-scale usage data sets from providers world-wide.

The purpose of the MESUR project is to study usage behavior in the scholarly process and therefore, usage modeling is a necessary component of the MESUR ontology. Given both usage and bibliographic data, it will be possible to generate and validate metrics for understanding the `value' of all types of scholarly artifacts. Currently, the scholarly community has one primary means of understanding the value of a journal and thus its authors: the ISI Impact Factor \cite{impactreivew:garfield1999}. With a semantic network data structure that includes not only article (and thus, journal) citation, but also authorship, usage, and institutional relationships, new metrics that not only rank journals, but also conferences, authors, and institutions will be created and validated.

Finally, the proposed ontology was engineered to handle an extremely large semantic network instantiation (on the order of 50 million articles with a corresponding 1 billion usage events). The MESUR ontology was engineered to make a distinction between required base-relationships and those, that if needed, can be inferred from the base-relations. Futhermore, due to the fact that the MESUR ontology was developed to support the large-scale analysis of usage, many of the metadata properties such as article title or author name are not explicitly represented in the ontology and thus, as will be demonstrated, such data can be accessed outside the triple store by reference to a relational database.

\section{Related Work}

Other efforts have produced and exploited scholarly ontologies, but they do not cover the needs of the MESUR project for two primary reasons.  First, they generally lack the integration of publication, citation and usage data, which MESUR requires in order to represent and analyze these crucial stages of the public scholarly communication process. Second, scalability appears to not have been a major concern when designing the ontologies and thus, instantiating them at the order of what MESUR will be representing is unfeasible. Sometimes, the ontology is too elaborate, adding complexity that rarely pays off for the simple reason that it is hard to realistically come by data to populate defined properties (e.g.~detailed author or affiliation information). Other times, the ontology requires the storage of information that cannot realistically be represented for vast data collections using current triple store technologies.

Several scholarly ontologies are available in the DAML Ontology Library\footnote{DAML Ontology Library available at: http://www.daml.org/ontologies/}. While they focus on bibliographic constructs, they do not model usage events.  The same is true of the Semantic Community Web Portal ontology \cite{semcomm:2000}, which, in addition maintains many detailed classes whose instantiation is unrealistic given what is recorded by modern scholarly information systems.  

The ScholOnto ontology was developed as part of an effort aimed at enabling researchers to describe and debate, via a semantic network, the contributions of a document, and its relationship to the literature \cite{scholonto:shum2000}.  While this ontology supports the concept of a scholarly document and a scholarly agent, it focuses on formally summarizing and interactively debating claims made in documents, not on expressing the actual use of documents.  Moreover, support for bibliographic data is minimal whereas support for discourse constructs, not required for MESUR, is very detailed.    

The ABC ontology \cite{abc:lagoze2001} was primarily engineered as a common conceptual model for the interoperability of a variety of metadata ontologies from different domains.  Although the ABC ontology is able to represent bibliographic and usage concepts by means of constructs such as artifact (e.g. article), agent (e.g. author), and action (e.g. use), it is designed at a level of generality that does not directly support the granularity required by the MESUR project.

An interesting ontology-based approach was developed by the Ingenta MetaStore project \cite{scholtrip:portwin2006}. Unfortunately, again, the Ingenta ontology does not support expressing usage of scholarly documents, which is a primary concern in MESUR. Nevertheless, the approach is inspiring because Ingenta faces significant challenges regarding scalability of the ontology-based representation, storage and access of their bibliographic metadata collection, which covers approximately 17 million journal articles.  However, the scale of the MESUR data set is several orders of magnitude larger, calling for optimizations wherever possible.  For example, given the MESUR project's focus on usage, storing bibliographic properties (author names, abstract, titles, etc.) in the triple store, as done by Ingenta, is not essential.  As a result, in order to improve triple store query efficiency, MESUR stores such data in a relational database, and the MESUR ontology does not explicitly represent these literals.

The principles espoused by the OntologyX\footnote{OntologyX available at: http://www.ontologyx.com/} ontology are inspiring. OntologyX uses \textit{context} classes as the ``glue" for relating other classes, an approach that was adopted for the MESUR ontology. For instance, the MESUR ontology does not have a direct relationship between an article and its publishing journal.  Instead, there exists a publishing context that serves as an N-ary operator uniting a journal, the article, its publication date, its authors, and auxiliary information such as the source of the bibliographic data. The context construct is intuitive and allows for future extensions to the ontology. OntologyX also helped to determine the primary abstract classes for the MESUR ontology. Unfortunately, OntologyX is a proprietary ontology for which very limited public information is available, making direct adoption unfeasible for MESUR. As a matter of fact, all inspiration was derived from a single PowerPoint presentation from the 2005 FBRB Workshop \cite{rust:ontx2005}.

Finally, in the realm of usage data representation, no ontology-based efforts were found. Nevertheless, the following existing schema-driven approaches were explored and served as inspiration: the OpenURL ContextObject approach to facilitate OAI-PMH-based harvesting of scholarly usage events \cite{bx:bollen2006}, the XML Log standard to represent digital library logs \cite{xmllog:goncalves2002}, and the COUNTER schema to express journal level usage statistics \cite{projec:shepherd2004}.

\section{Leveraging Relational Database Technology \label{sec:leverage}}

The MESUR project makes use of a triple store to represent and access its collected data. While the triple store is still a maturing technology, it provides many advantages over the relational database model. For one, the network-based representation supports the use of network analysis algorithms. For the purposes of the MESUR project, a network-based approach to data analysis will play a major role in quantifying the value of the scholarly artifacts contained within it. Other benefits that are found with triple store technologies that are not easily reproducible within the relational database framework include ease of schema extension and ontological inferencing.

A novel contribution of the presented ontology is its solution to the problem of scalability found in modern triple store technologies \cite{lee:triple2004}. While semantic networks provide a flexible medium for representing and searching knowledge, current triple store applications do not support the amount of data that can be represented at the upper limit of what is possible with modern relational database technologies. Therefore, it was necessary to be selective of what information is actually modeled by the MESUR ontology. For the MESUR project, much of the data associated with each scholarly artifact is maintained outside the triple store in a relational database.
  
The typical bibliographic record contains, for example, an article's identifiers (e.g.~DOI, SICI, etc.), authors, title, journal/conference/book, volume, issue, number, and page numbers. Typical usage information contains, for example, the users identifier (e.g.~IP address), the time of the usage event, and a session identifier. An example of the various bibliographic and usage properties are outlined in the Table \ref{tab:docprop} and Table \ref{tab:useprop}, respectively. Note that the connection between the bibliographic record and the usage event occurs through the doc\_id (bolded properties). The doc\_id is a internally generated identifier created during the MESUR project's ingestion process.

\begin{table}[htdp]
\begin{scriptsize}
\begin{center}
\begin{tabular}{|l|l|}\hline
\textbf{property} & \textbf{value} \\\hline\hline
title             & The Convergence of Digital Libraries ...  \\\hline
author(s)   & Rodriguez, Bollen, Van de Sompel \\\hline
collection  & Journal of Information Science \\\hline
publisher  & Sage Publications \\\hline 
date           & 2006 \\\hline
start page & 149  \\\hline
end page & 159  \\\hline
volume     & 32 \\\hline
issue         & 2 \\\hline
doi  		& 10.1177/0165551506062327 \\\hline
doc\_id 	& \textbf{b5e1ab73-26b5-41f0-a83f-b47b4d737} \\\hline
\end{tabular}
\end{center}
\caption{Example bibliographic properties \label{tab:docprop}}
\end{scriptsize}
\end{table}

\begin{table}[htdp]
\begin{scriptsize}
\begin{center}
\begin{tabular}{|l|l|}\hline
\textbf{property} & \textbf{value} \\\hline\hline
event\_id          & 45563ac2-c7d4-4669-ab9c-ac5129535ee5 \\\hline
time                   & 2006-09-27 00:00:03 \\\hline
agent               & 4AD2FD457EB59CE08AAAF6EA2A63F \\\hline
session            & C3044206 \\\hline
affiliation          & California State University, Los Angeles \\\hline
doc\_id         & \textbf{b5e1ab73-26b5-41f0-a83f-b47b4d737} \\\hline
\end{tabular}
\end{center}
\caption{Example usage properites \label{tab:useprop}}
\end{scriptsize}
\end{table}

\begin{figure}[h!]
	\centering
	\includegraphics[width=0.375\textwidth]{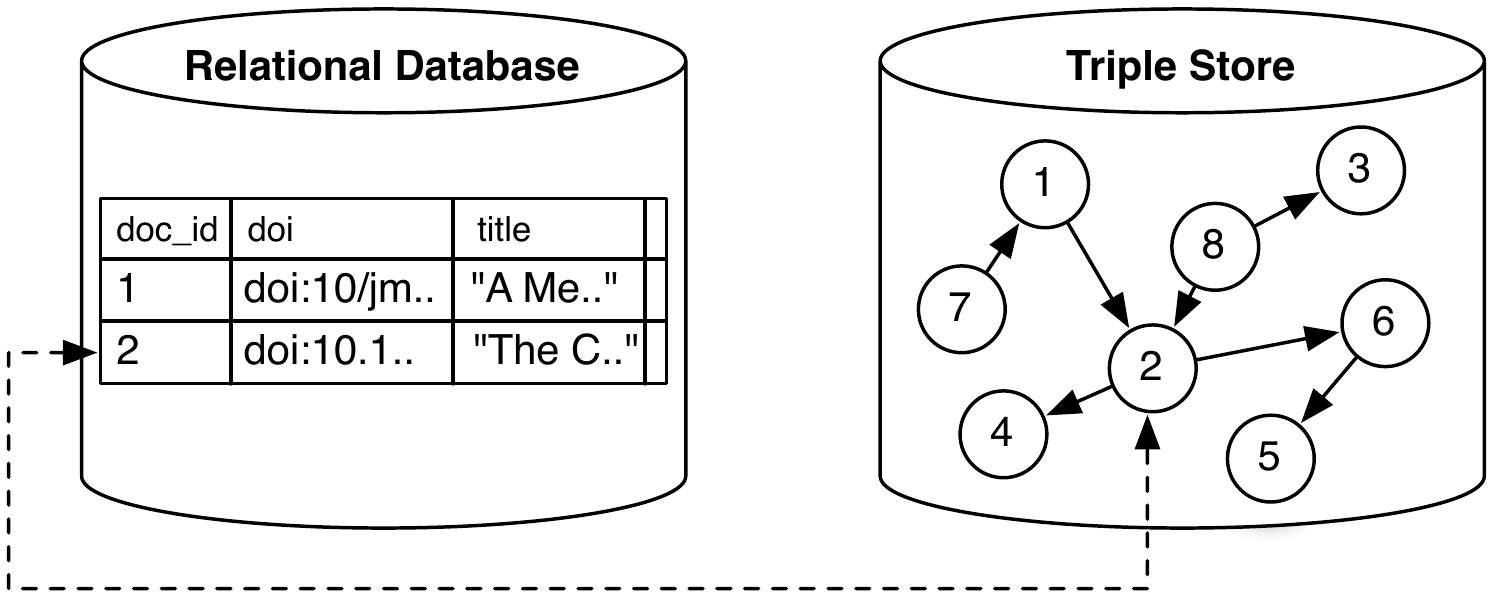}
	 \caption{The relationship between the relational database and the triple store \label{fig:query-model}}
\end{figure}

The two tables demonstrate how bibliographic and usage data can be easily represented in a relational database. From the relational database representation, a RDF N-Triple\footnote{N-Triple available at: \\ http://www.w3.org/2001/sw/RDFCore/ntriples/} data file can be generated. One such solution for this relational database to triple store mapping is the D2R mapper \cite{d2r:bizer2003}. However, note that not all data in the relational database is exported to this intermediate format. Instead, only those properties that promote triple store scalability and usage research were included. Thus, article titles, journal issues and volumes, names of authors, to name a few, are not explicitly represented within the triple store and thus, are not modeled by the ontology. If a particular artifact property that is not in the ontology is required for a computation, the computing algorithm references the relational database holding the complete representation the acquired data. For example, bi-directional resolution of the artifact with doc\_id $2$ is depicted in Figure \ref{fig:query-model} where the resolving identifier is specific to the artifact (for the sake of diagram readability, assume that $2$ is b5e1ab73-26b5-41f0-a83f-b47b4d737 from Table \ref{tab:docprop} and \ref{tab:useprop}). This model is counter to what is seen in other scholarly ontologies such as the Ingenta ontology \cite{scholtrip:portwin2006}. This design choice was a major factor that prompted the engineering of a new ontology for bibliographic and usage modeling.

\section{The MESUR Ontology \label{sec:mesuront}}

The MESUR ontology is currently at version 2007-01 at \texttt{http://www.mesur.org/schemas/2007-01/mesur} (abbreviated \texttt{mesur}). Full HTML documentation of the ontology can be found at the namespace URI. The following sections will describe how bibliographic and usage data is modeled to meet the requirements of understanding large-scale usage behavior, while at the same time promoting scalability.

\subsection{The Primary Classes}

\begin{figure}[h!]
	\centering
	\includegraphics[width=0.25\textwidth]{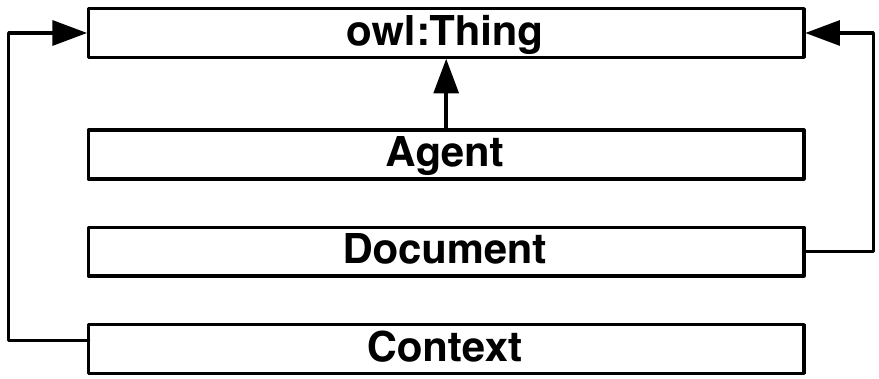}
	 \caption{The primary classes of the MESUR ontology \label{fig:uml-thing}}
\end{figure}

The most general class in OWL is \texttt{owl:Thing}. The MESUR ontology provides three subclasses of \texttt{owl:Thing}. These MESUR classes are \texttt{mesur:Agent}, \texttt{mesur:Document}, and \texttt{mesur:Context}\footnote{For the remainder of this article, all classes that are not explicitly namespaced are from the \texttt{mesur} namespace.}. This is represented in Figure \ref{fig:uml-thing} where an edge denotes a \texttt{rdfs:subClassOf} relationship.

The \texttt{Context} classes serve as the ``glue" by which \texttt{Agent}s and \texttt{Document}s interact. A \texttt{Context} is analogous to \texttt{rdf:Bag} in that it is an N-ary operator unifying the literals and objects pointed to by its respective properties. All relationships between \texttt{Agent}s and \texttt{Document}s occurs through a particular \texttt{Context}. However, as will be demonstrated, direct relationships can be inferred.  All inferred properties are denoted by the ``(i)" notation in the following UML class diagrams.  All inferred properties are superfluous relationships since there is no loss of information by excluding their instantiation (the information is contained in other relationships). The algorithms for inferring them will be discussed in their respective \texttt{Context} subsection.

\begin{figure}[h!]
	\centering
	\includegraphics[width=0.475\textwidth]{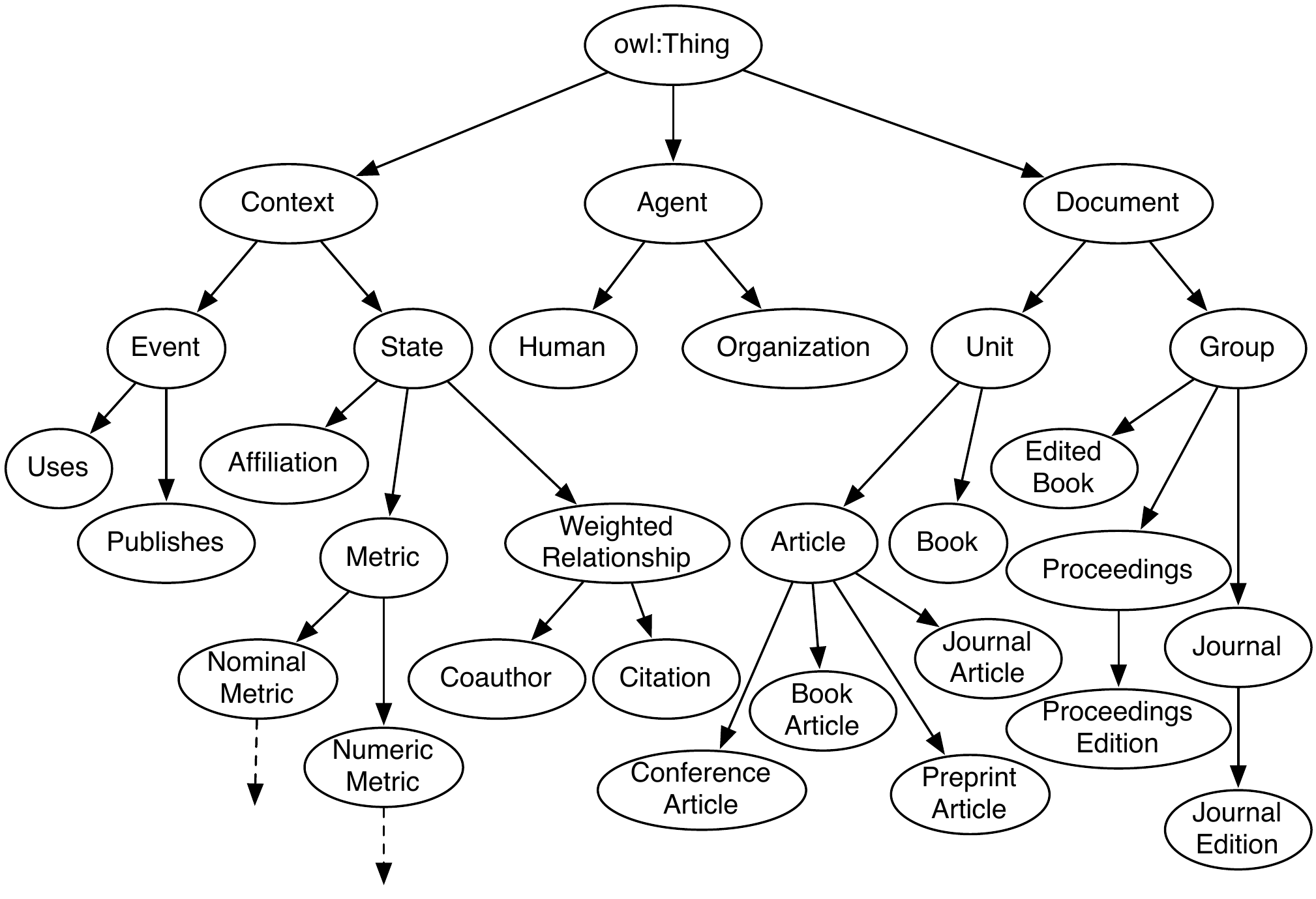}
	 \caption{MESUR taxonomy \label{fig:full-ontology}}
\end{figure}

Currently, all the MESUR classes are specifications or generalizations of other classes. No holonymy/meronymy (composite) class definitions are used at this stage of the ontology's development. Figure \ref{fig:full-ontology} presents the complete taxonomy of the MESUR ontology. This diagram primarily serves as a reference. Each class will be discussed in the following sections.

\subsection{The Agent Classes}

\begin{figure}[h!]
	\centering
	\includegraphics[width=0.475\textwidth]{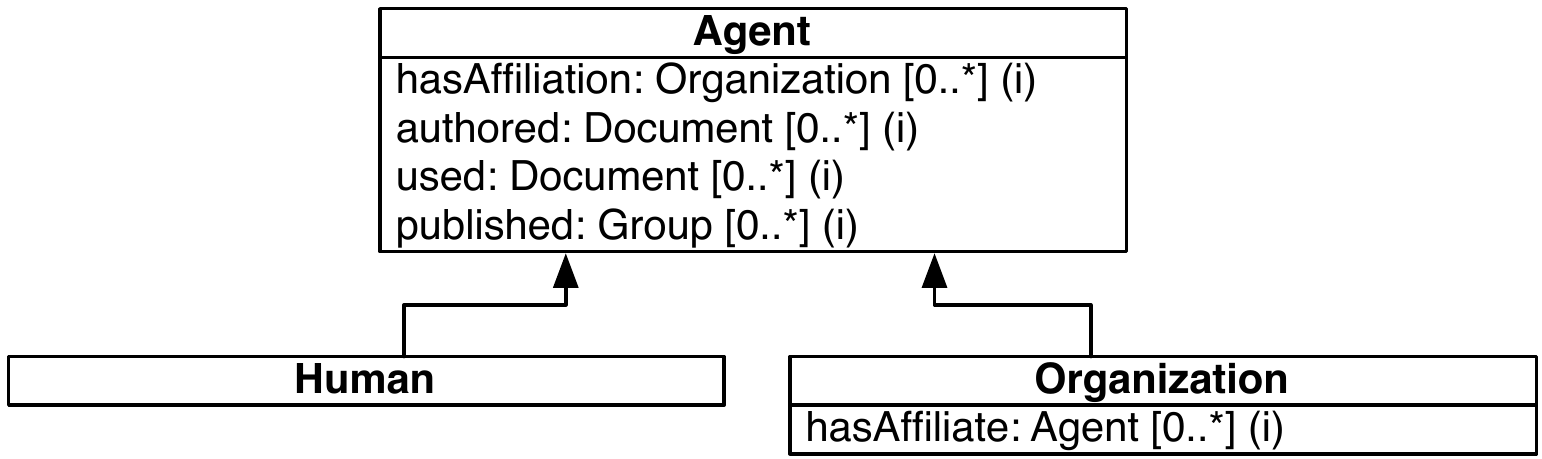}
	 \caption{Classes of Agent and their properties \label{fig:uml-agent}}
\end{figure}

The \texttt{Agent} taxonomy is diagrammed in Figure \ref{fig:uml-agent}. An \texttt{Agent} can either be a \texttt{Human} or an \texttt{Organization}. A \texttt{Human} is an actual individual whether that individual can be uniquely identified (e.g.~an document author) or not (e.g.~a document user). The \texttt{authored} property is an inferred relationship and denotes that an \texttt{Agent} authored a particular \texttt{Document} and the \texttt{published} property denotes that an \texttt{Agent} has published a \texttt{Document}. The \texttt{authored} and \texttt{published} property can be inferred by information within the \texttt{Publishes} context discussed later. Similarly, the \texttt{used} property denotes that an \texttt{Agent} has used a particular \texttt{Document}. The \texttt{used} property can be inferred from the \texttt{Uses} context.

An \texttt{Organization} is a class that is used for both bibliographic and usage provenance purposes. Given that bibliographic and usage data, at the large-scale, must be harvested from multiple institutions, it is necessary to make a distinction between the various data providers. In many cases, an \texttt{Organization} can be both a bibliographic (e.g.~a publisher) and a usage (e.g.~a repository) provider. Furthermore, an \texttt{Organization} can also be an author's academic institution (e.g.~a university).

Finally, all \texttt{Agent}s can have any number of affiliations. For an \texttt{Organization}, this is a recursive definition which allows \texttt an {Organization} to have many affiliate \texttt{Organization}s while at the same time allowing for the \texttt{Human} leaf nodes of an \texttt{Organization} to be represented by the same construct. The rules governing the inference of the \texttt{hasAffiliation} and \texttt{hasAffiliate} properties are discussed in the section describing the \texttt{Affiliation} context.

 \subsection{The Document Classes}
 
\begin{figure}[h!]
	\centering
	\includegraphics[width=0.475\textwidth]{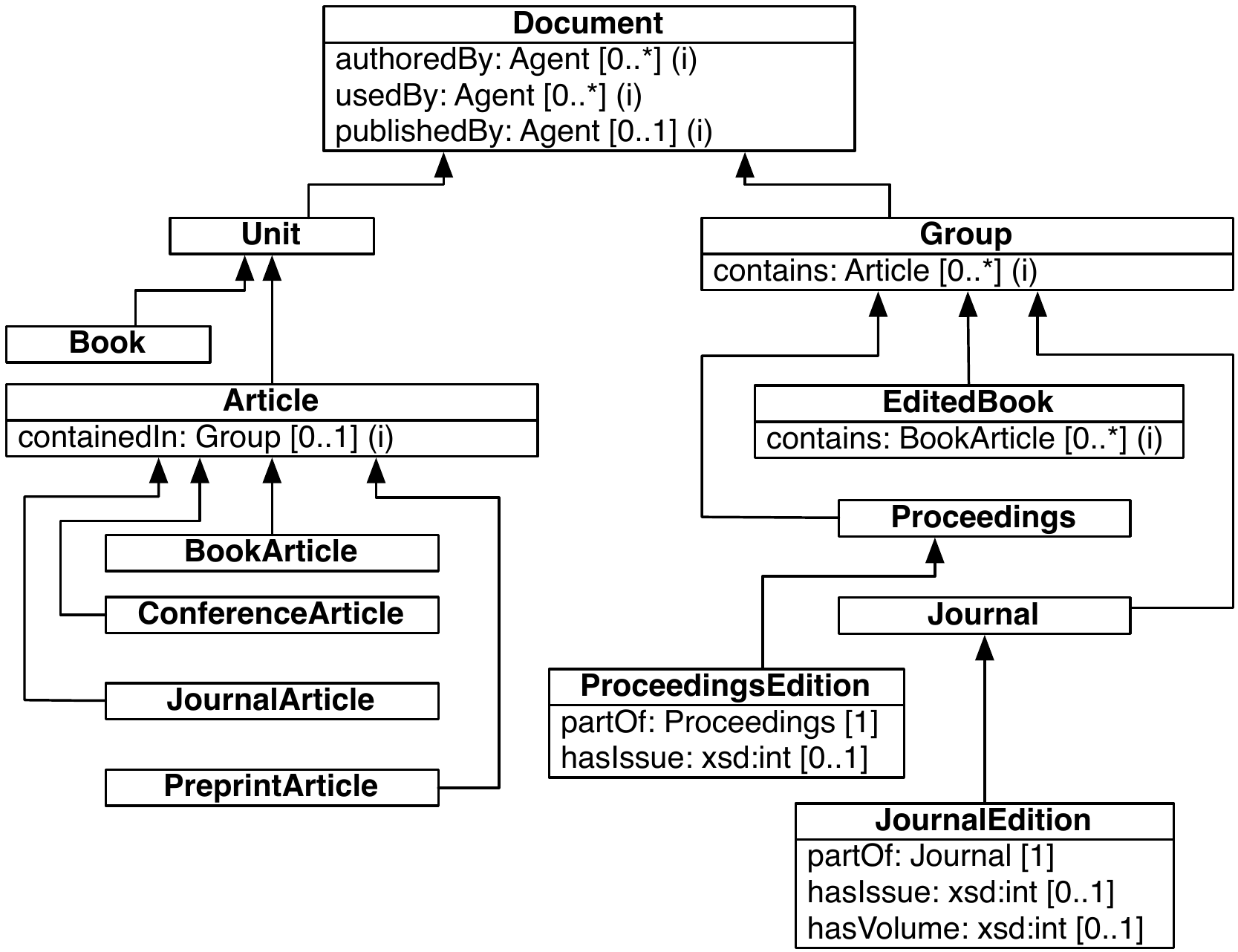}
	 \caption{Classes of Document and their properties \label{fig:uml-document} }
\end{figure}

A \texttt{Document} is an abstract concept of a particular scholarly product such as those depicted in Figure \ref{fig:uml-document}.

In general, \texttt{Document} objects are those artifacts that are written, used, and published by \texttt{Agent}s. Thus, a \texttt{Document} can be a specific article, a book, or some grouping such as a \texttt{Journal}, conference \texttt{Proceedings}, or an \texttt{EditedBook}. There are two \texttt{Document} subclasses to denote whether the \texttt{Document} is a collection (\texttt{Group}) or an individually written work (\texttt{Unit}). A \texttt{Journal} and \texttt{Proceedings} is an abstract concept of a collection of volumes/issues. An edition to a proceedings or journal is associated with its abstract \texttt{Group} by the \texttt{partOf} property. The \texttt{authoredBy}, \texttt{containedIn}, \texttt{publishedBy}, and \texttt{contains} properties can be inferred from the \texttt{Publishes} context. Also, the \texttt{usedBy} property can be inferred from the \texttt{Uses} context.

\subsection{The Context Classes}

\begin{figure}[h!]
	\centering
	\includegraphics[width=0.325\textwidth]{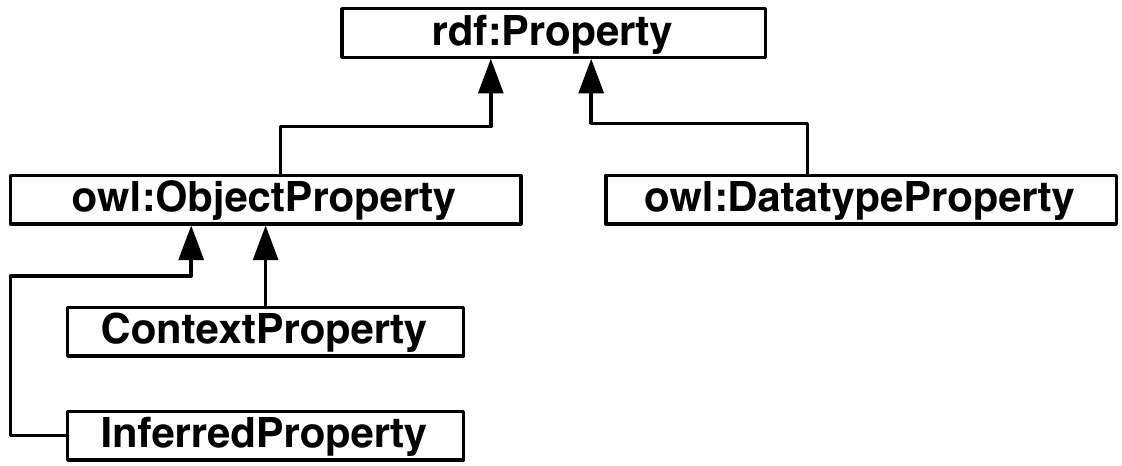}
	 \caption{The abstract MESUR property classes \label{fig:uml-property}}
\end{figure}

As previously stated, all properties from the \texttt{Agent} and \texttt{Document} classes that are marked by the ``(i)" notation are inferred properties. These properties can be automatically generated by inference algorithms and thus, are not required for insertion into the triple store. What this means is that inherent in the triple store is the data necessary to infer such relationships. Depending on the time (e.g.~query complexity) and space (e.g.~disk space allocation) constraints, the inclusion of these inferred properties is determined. At any time, these properties can be inserted or removed from the triple store. The various inferred properties are determined from their respective \texttt{Context} objects. Therefore, the MESUR \texttt{owl:ObjectProperty} taxonomy provides two types of object properties: \texttt{ContextProperty} and \texttt{InferredProperty} (see Figure \ref{fig:uml-property}).

\begin{figure}[h!]
	\centering
	\includegraphics[width=0.475\textwidth]{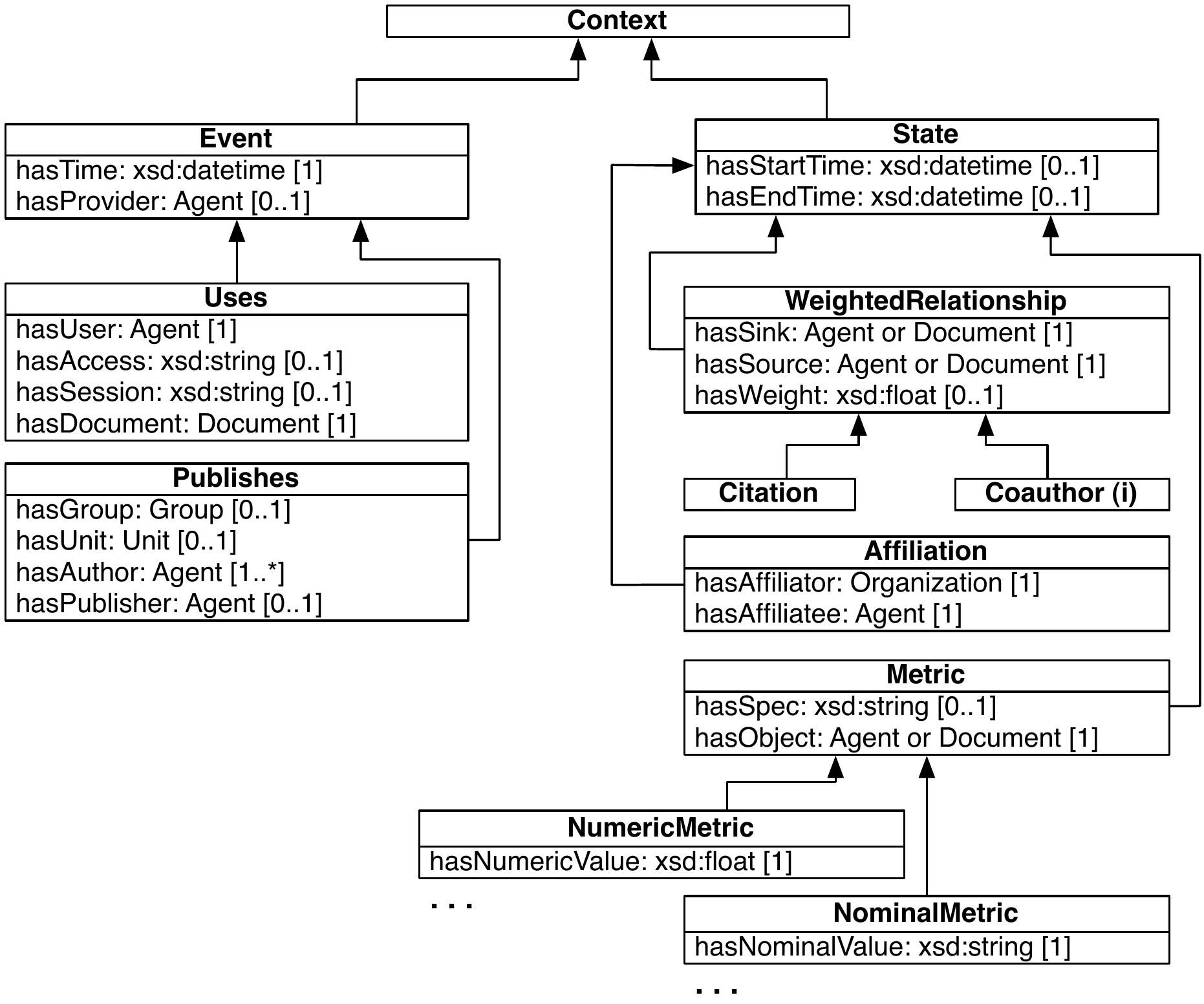}
	 \caption{Classes of Context and their properties \label{fig:uml-context}}
\end{figure}

 A \texttt{Context} class is an N-ary operator much like an \texttt{rdf:Bag}. Current triple store technology expresses tertiary relationships. That means that only three resources are related by a semantic network edge (i.e.~a subject URI, predicate URI, and object URI). However, many real-world relationships are the product of multiple interacting objects. It is the role of the various \texttt{Context} classes to provide relationships for more than three URIs. The \texttt{Context} classes are represented in Figure \ref{fig:uml-context}.

The \texttt{Context} class has two subclasses: \texttt{Event} and \texttt{State}. An \texttt{Event} is some measurement done by some provider at a particular point in time. For example, the \texttt{Publishes} and \texttt{Uses} events are recorded by publisher and repositories at some point in time. As a side note, the \texttt{hasProvider} property of the \texttt{Event} class is an efficient model for the representation of provenance constructs. Instead of reifying every statement with provenance data (e.g.~triple $x$ was supplied by provider $y$ \cite{scholtrip:portwin2006}), a single triple is provided for each \texttt{Event} (e.g.~event $x$ was supplied by provider $y$). 

On the other side of the \texttt{Context} taxonomy are the \texttt{State} contexts. A \texttt{State} is some measurement that can, in some cases, occur over a span of time and are used to represent complex relationships between artifacts or as a way of attaching high-level properties (i.e.~metadata) to an artifact. The next sections will provide a detailed description of each \texttt{Context} class along with SPAQRL queries for inferring all the aforementioned \texttt{InferredProperty} properties.
 
 \subsubsection{The Publishes Context}

\begin{figure}[h!]
	\centering
	\includegraphics[width=0.475\textwidth]{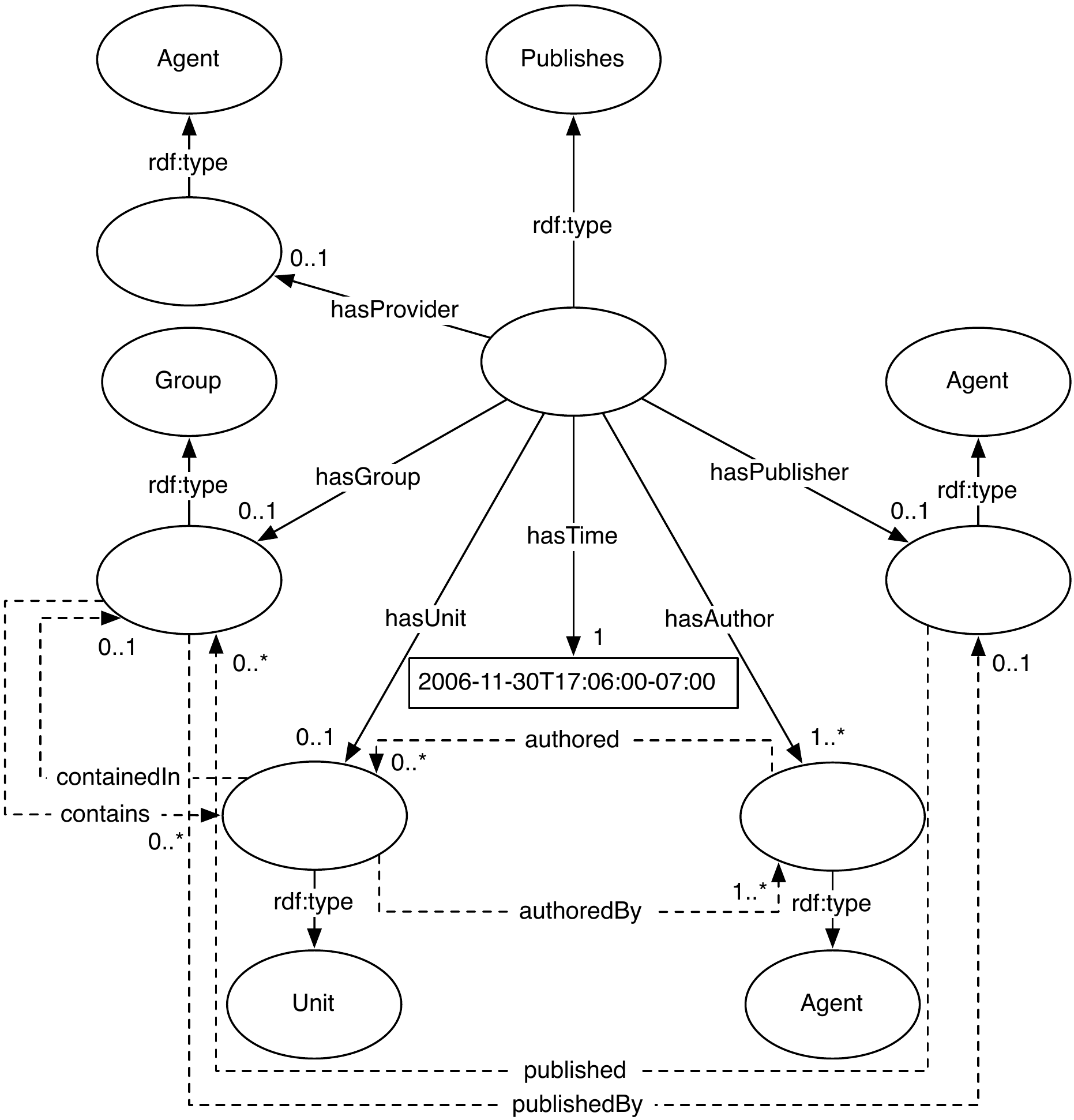}
	 \caption{Example Publishes Context \label{fig:publish-context}}
\end{figure}

A \texttt{Publishes} event states, in words, that a particular bibliographic data provider has acknowledged that a set of authors have authored a unit that was published in a group by some publisher at a particular point in time. A \texttt{Publishes} object relates a single bibliographic data provider, \texttt{Agent} authors, a \texttt{Unit}, an \texttt{Agent} publisher, a \texttt{Group}, and a publication ISO-8601 date time literal\footnote{ISO-8601 available at: http://www.w3.org/TR/NOTE-datetime/}. Figure \ref{fig:publish-context} represents a \texttt{Publishes} context and the inferable properties (dashed edges) of the various associated artifacts. All inferred properties have a respective inverse relationship. Note that both \texttt{PreprintArticle} and \texttt{Book} publishing are represented with OWL restrictions (i.e.~they are not published in a \texttt{Group}). The details of these restrictions can be found in the actual ontology definition.

The dashed edges in Figure \ref{fig:publish-context} denote properties that are a \texttt{rdfs:subClassOf} the \texttt{InferredProperty}. For instance, the abstract triple $\langle \texttt{Author}, \texttt{authors}, \texttt{Document} \rangle$ is inferred given the results of the following SPARQL query, where for the sake of brevity, the \texttt{PREFIX} declarations are removed and the \texttt{INSERT} statement represents the insert of its triple argument into the triple store\footnote{Please note that all the presented SPARQL queries are not optimized for speed, but instead, are optimized for readability.}.

\begin{scriptsize}
\begin{lstlisting}
SELECT  ?a ?b
WHERE   
	( ?x rdf:type mesur:Publishes ) 
	( ?x mesur:hasUnit ?a )
	( ?x mesur:hasAuthor ?b )

INSERT < ?a mesur:authoredBy ?b >	
INSERT < ?b mesur:authored ?a > .
\end{lstlisting}
\end{scriptsize}

To infer the \texttt{Group} property \texttt{contains} and \texttt{Unit} property \texttt{containedIn}, the following SPARQL query and \texttt{INSERT} statements suffice.

\begin{scriptsize}
\begin{lstlisting}
SELECT  ?a ?b
WHERE   
	( ?x rdf:type mesur:Publishes ) 
	( ?x mesur:hasUnit ?a )
	( ?x mesur:hasGroup ?b )

INSERT < ?a mesur:containedIn ?b >
INSERT < ?b mesur:contains ?a > .
\end{lstlisting}
\end{scriptsize}

Finally, the \texttt{published} and \texttt{publishedBy} properties are inferred by:

\begin{scriptsize}
\begin{lstlisting}
SELECT  ?a ?b
WHERE   
	( ?x rdf:type mesur:Publishes ) 
	( ?x mesur:hasPublisher ?a )
	( ?x mesur:hasGroup ?b )
	
INSERT < ?a mesur:published ?b >
INSERT < ?b mesur:publishedBy ?a > .
\end{lstlisting}
\end{scriptsize}

\subsubsection{The Uses Context}

\begin{figure}[h!]
	\centering
	\includegraphics[width=0.45\textwidth]{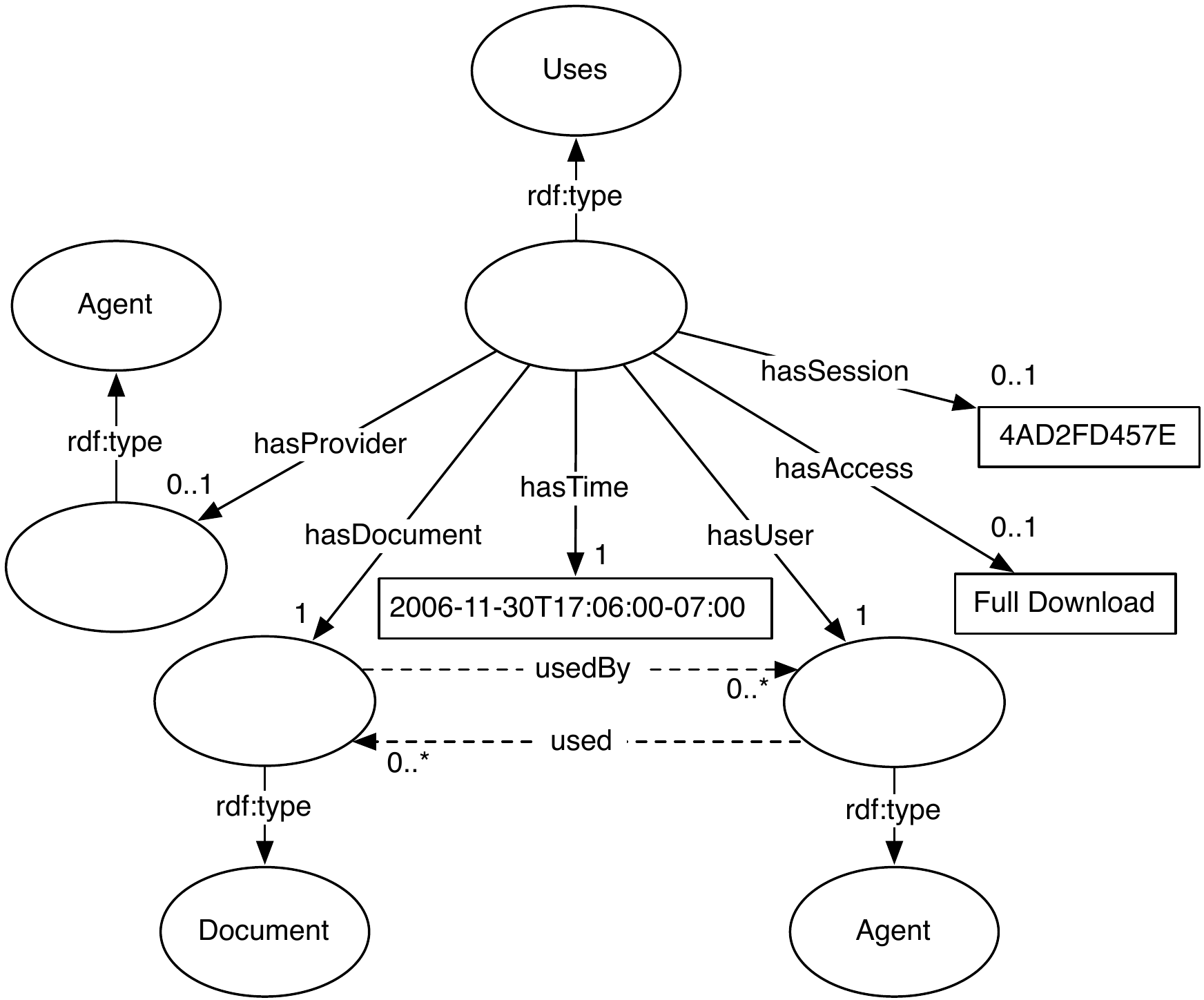}
	 \caption{Example Uses Context \label{fig:use-context}}
\end{figure}

The \texttt{Uses} context denotes a single usage event where an \texttt{Agent} uses a \texttt{Document} at a particular point in time. The \texttt{Uses} context is diagrammed in Figure \ref{fig:use-context}.  Like the \texttt{Publishes} context, the \texttt{Uses} context is an N-ary construct. Depending on the usage provider, a session identifier and access type is recorded. A session identifier denotes the user's login session. An access type denotes, for example, whether the used \texttt{Document} had its abstract viewed or was fully downloaded.

The following SPARQL query and \texttt{INSERT} statements represent the inference of the \texttt{usedBy} and \texttt{used} inverse properties of an \texttt{Article} document and \texttt{Agent}, respectively. Also, note the last two \texttt{INSERT} statements. These statements demonstrate how \texttt{Group} usage information can also be inferred.

\begin{scriptsize}
\begin{lstlisting}
SELECT  ?a ?b ?c
WHERE   
	( ?x rdf:type mesur:Uses ) 
	( ?x mesur:hasDocument ?a )
	( ?a rdf:type mesur:Article )
	( ?x mesur:hasUser ?b )
	( ?y rdf:type mesur:Publishes )
	( ?y mesur:hasUnit ?a )
	( ?y mesur:hasGroup ?c )
	
INSERT < ?a mesur:usedBy ?b >
INSERT < ?b mesur:used ?a >
INSERT < ?c mesur:usedBy ?b >
INSERT < ?b mesur:used ?c > .
\end{lstlisting}
\end{scriptsize}

\subsubsection{The Weighted Relationship Context}

In many instances, one artifact is related to another by a particular semantic. However, in some instance, one artifact is related to another by a semantic label and a floating point weight value. Furthermore, that weighted relationship may have been recorded over some period of time. The \texttt{WeightedRelationship} state context is used to represent such relationships. 

\begin{figure}[h!]
	\centering
	\includegraphics[width=0.475\textwidth]{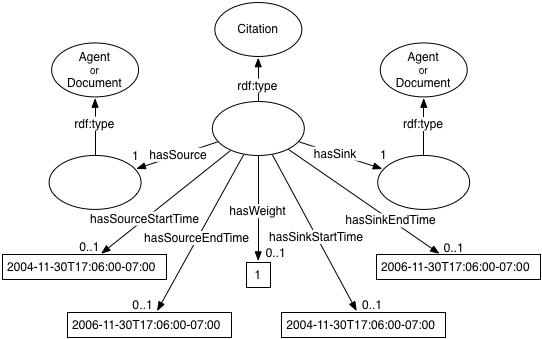}
	 \caption{Example Citation Context \label{fig:citation-context}}
\end{figure}

The \texttt{Citation} state context denotes a weighted citation and is a \texttt{rdfs:subClassOf} the \texttt{WeightedRelationship}. For \texttt{Unit} to \texttt{Unit} citation, the weight value is $1.0$ (or no weight property to reduce the triple store footprint) and there are no start and end time points. However, for \texttt{Group} to \texttt{Group} citations, the weight of the \texttt{Citation} represents how many times a particular \texttt{Group} cites another over some period of time. Hence, it is necessary to denote the start and end points of both the source and the sink nodes. Figure \ref{fig:citation-context} diagrams a \texttt{Citation} context. Furthermore, the sink and source types can be either an \texttt{Agent} or a \texttt{Document}, thus, \texttt{Organization} to \texttt{Organization} citations can be represented.

Given \texttt{Unit} to \texttt{Unit} citations, the \texttt{Citation} weight between any two \texttt{Group}s can be inferred. The following example SPARQL query generates the \texttt{Citation} object for citations from 2007 articles in the Journal of Informetrics (ISSN: 1751-1577) to 2005-2006 articles in Scientometrics (ISSN: 0138-9130). Assume that the URI of the journals are their ISSN numbers, the date time is represented as a year instead of the lengthy ISO-8601 representation, and the \texttt{COUNT} command is analogous to the SQL \texttt{COUNT} command (i.e.~returns the number of elements returned by the variable binding).

\begin{scriptsize}
\begin{lstlisting}
SELECT  ?x
WHERE 
	( ?x rdf:type mesur:Citation )
	( ?x mesur:hasSource ?a )
	( ?x mesur:hasSink ?b )
	( ?a rdf:type mesur:Article )
	( ?b rdf:type mesur:Article )
	( ?y rdf:type mesur:Publishes )
	( ?z rdf:type mesur:Publishes )
	( ?y mesur:hasTime ?t) 
		AND (?t > 2004 AND ?t < 2007)
	( ?z mesur:hasTime ?u) AND ?u = 2007
	( ?y mesur:hasUnit ?a )
	( ?z mesur:hasUnit ?b )
	( ?y mesur:hasGroup ?c )
	( ?z mesur:hasGroup ?d )
	( ?c mesur:partOf urn:issn:1751-1577 )
	( ?d mesur:partOf urn:issn:0138-9130 )

INSERT < _123 rdf:type mesur:Citation >
INSERT < _123 mesur:hasSource urn:issn:1751-1577 >
INSERT < _123 mesur:hasSink urn:issn:0138-9130 >
INSERT < _123 mesur:hasWeight COUNT(?x) >
INSERT < _123 mesur.hasSourceStartTime 2007 >
INSERT < _123 mesur:hasSourceEndTime 2007 > 
INSERT < _123 mesur.hasSinkStartTime 2005 >
INSERT < _123 mesur:hasSinkEndTime 2006 > .
\end{lstlisting}
\end{scriptsize}

\begin{figure}[h!]
	\centering
	\includegraphics[width=0.375\textwidth]{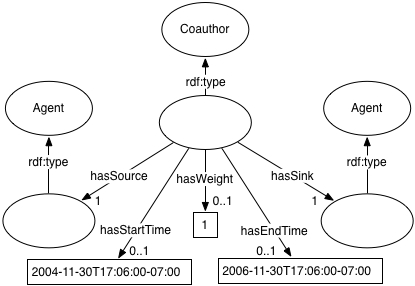}
	 \caption{Example Coauthor Context \label{fig:coauthor-context}}
\end{figure}

Figure \ref{fig:coauthor-context} diagrams the \texttt{Coauthor} weighted relationship context. The weight value of this relationship denotes the number of times two authors have coauthored together over a some period of time.

The following SPARQL query demonstrates how to infer the weighted \texttt{Coauthor} relationship between the authors Marko (\texttt{lanl:marko}) and Herbert (\texttt{lanl:herbertv}) over all time. A time period for coauthorship counting can be inserted in a fashion similar to the \texttt{Citation} example previous.

\begin{scriptsize}
\begin{lstlisting}
SELECT ?x
WHERE 
	( ?x rdf:type mesur:Publishes ) 
	( ?x mesur:hasAuthor lanl:marko )
	( ?x mesur:hasAuthor lanl:herbertv ) 

INSERT < _123 rdf:type mesur:Coauthor >
INSERT < _123 mesur:hasSource lanl:marko >
INSERT < _123 mesur:hasSink lanl:herbertv >
INSERT < _123 mesur:hasWeight COUNT(?x) >
INSERT < _456 rdf:type mesur:Coauthor >
INSERT < _456 mesur:hasSource lanl:herbertv >
INSERT < _456 mesur:hasSink lanl:marko >
INSERT < _456 mesur:hasWeight COUNT(?x) > .
\end{lstlisting}
\end{scriptsize}

\subsubsection{The Affiliation Context}

\begin{figure}[h!]
	\centering
	\includegraphics[width=0.375\textwidth]{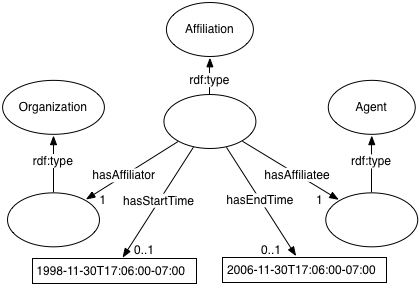}
	 \caption{Example Affiliation Context \label{fig:affiliation-context}}
\end{figure}

An \texttt{Affiliation} context denotes that a particular \texttt{Human} is affiliated with an \texttt{Organization} or that an \texttt{Organization} is affiliated with another \texttt{Organization}. An \texttt{Affiliation} can be represented as occurring over a particular period of time. An example of an \texttt{Affiliation} state context is diagrammed in Figure \ref{fig:affiliation-context}.

The \texttt{hasAffiliate} and \texttt{hasAffiliation} properties of the \texttt{Agent} classes can be inferred by the following SPARQL query.

\begin{scriptsize}
\begin{lstlisting}
SELECT  ?a ?b
WHERE   
	( ?x rdf:type mesur:Affiliation ) 
	( ?x mesur:hasAffiliator ?a )
	( ?x mesur:hasAffiliatee ?b )
	
INSERT < ?a mesur:hasAffiliate?b >
INSERT < ?b mesur:hasAffiliation?a > .
\end{lstlisting}
\end{scriptsize}

\subsubsection{The Metric Context}

\begin{figure}[h!]
	\centering
	\includegraphics[width=0.375\textwidth]{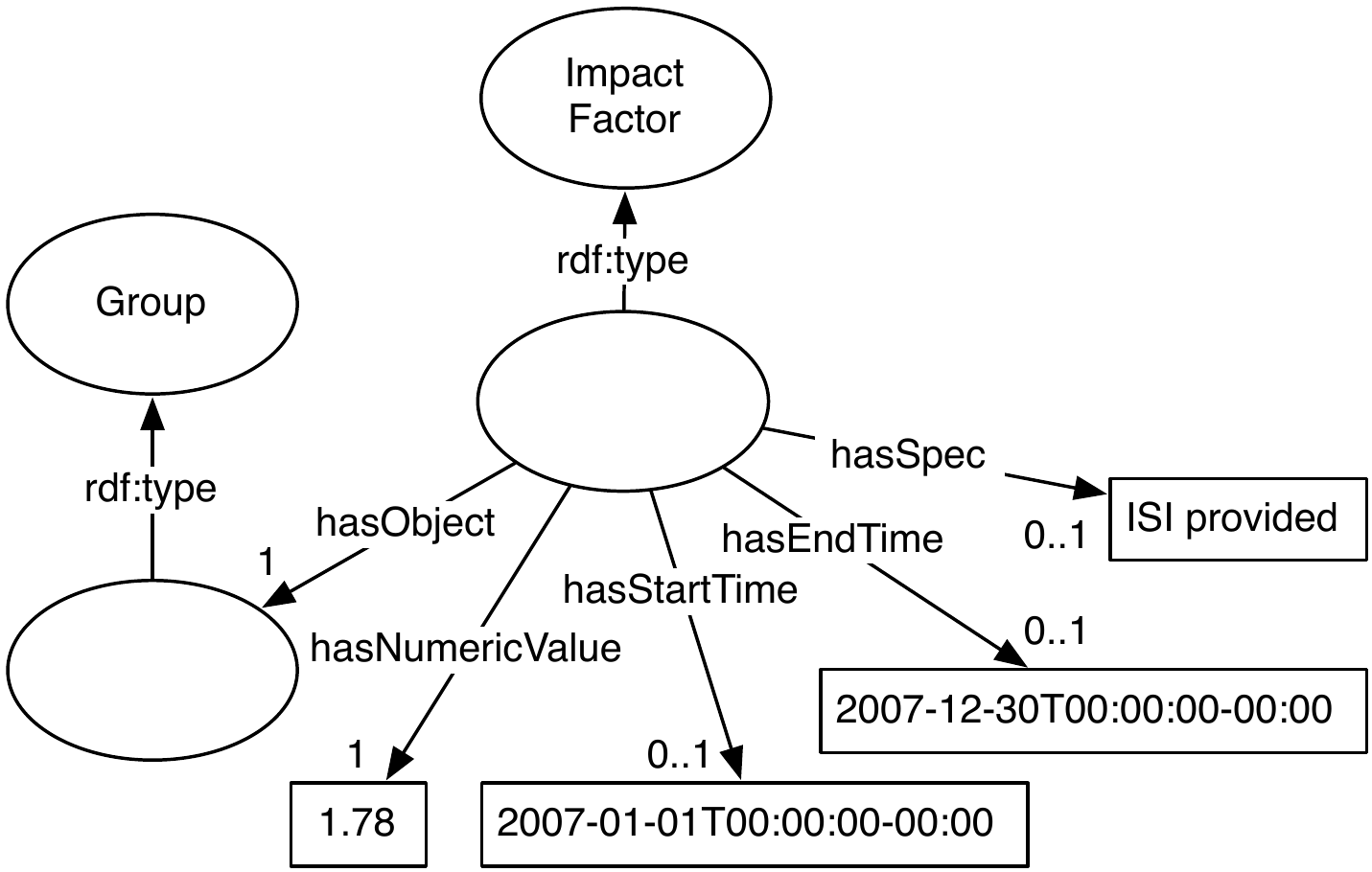}
	 \caption{Example Impact Factor Context \label{fig:numeric-context}}
\end{figure}

The primary objective of the MESUR project is to study the relationship between usage-based value metrics (e.g.~Usage Impact Factor \cite{usage:bollen2006}) and citation-based value metrics (e.g.~ISI Impact Factor \cite{impactreivew:garfield1999} and the Y-Factor \cite{journalstatus:bollen2006}). The \texttt{Metric} context allows for the explicit representation of such metrics. The \texttt{Metric} context has both the \texttt{NumericMetric} and \texttt{NominalMetric} subclasses. Figure \ref{fig:numeric-context} diagrams the 2007 \texttt{ImpactFactor} numeric metric context for a \texttt{Group}. Note that the \texttt{Context} hierarchy in Figure \ref{fig:uml-context} does not represent the set of \texttt{Metric}s explored by the MESUR project. This taxonomy will be presented in a future publication.

The example SPARQL query and respective \texttt{INSERT} statements demonstrate how to calculate the 2007 Impact Factor for the Proceedings of the Joint Conference on Digital Libraries (JCDL ISSN: 1082-9873). The 2007 Impact Factor for the JCDL is defined as the number of citations from any \texttt{Unit} published in 2007 to articles in the JCDL proceedings published in either 2005 or 2006 normalized by the total number of articles published by JCDL in 2005 and 2006 \cite{impactreivew:garfield1999}.

\begin{scriptsize}
\begin{lstlisting}
SELECT  ?x
WHERE   
	( ?x rdf:type mesur:Publishes ) 
	( ?x mesur:hasUnit ?a )
	( ?x mesur:hasGroup ?b )
	( ?b mesur:partOf urn:issn:1082-9873 )
	( ?x mesur:hasTime ?t ) AND 
		(?t > 2004 AND ?t < 2007)
	( ?y rdf:type mesur:Citation )
	( ?y mesur:hasSource ?c )
	( ?y mesur:hasSink ?a )
	( ?z rdf:type mesur:Publishes )
	( ?z mesur:hasUnit ?c )
	( ?z mesur:hasTime ?u) AND ?u = 2007
	
SELECT  ?y
WHERE   
	( ?y rdf:type mesur:Publishes )
	( ?y mesur:hasGroup ?a )
	( ?a mesur:partOf urn:issn:1082-9873 )
	( ?y mesur:hasTime ?t ) AND 
		(?t > 2004 AND ?t < 2007)

INSERT < _123 rdf:type mesur:ImpactFactor >
INSERT < _123 mesur:hasObject urn:issn:1082-9873 >
INSERT < _123 mesur:hasStartTime 2007 >
INSERT < _123 mesur:hasEndTime 2007 >
INSERT < _123 mesur:hasNumbericValue 
			(COUNT(?x) / COUNT(?y)) > .
\end{lstlisting}
\end{scriptsize}

The 2007 Usage Impact Factor for the JCDL Proceedings can be calculated by using the following SPARQL queries and \texttt{INSERT} commands. The 2007 Usage Impact Factor for the JCDL is defined as the number of usage events in 2007 that pertain to articles published in the JCDL proceedings in either 2005 or 2006 normalized by the total number of articles published by the JCDL in 2005 and 2006 \cite{usage:bollen2006}.

\begin{scriptsize}
\begin{lstlisting}
SELECT  ?x
WHERE   
	( ?x rdf:type mesur:Uses ) 
	( ?x mesur:hasDocument ?a )
	( ?x mesur:hasTime ?t ) AND ?t = 2007
	( ?y rdf:type mesur:Publishes )
	( ?y mesur:hasUnit ?a )
	( ?y mesur:hasGroup ?c )
	( ?c mesur:partOf urn:issn:1082-9873 ) 
	( ?y mesur:hasTime ?u ) AND 
		(?u > 2004 AND ?u < 2007)

SELECT  ?y
WHERE   
	( ?y rdf:type mesur:Publishes )
	( ?y mesur:hasGroup ?a )
	( ?a mesur:partOf urn:issn:1082-9873 )
	( ?y mesur:hasTime ?t ) AND 
		(?t > 2004 OR ?t < 2007)

INSERT < _123 rdf:type mesur:UsageImpactFactor >
INSERT < _123 mesur:hasObject urn:issn:1082-9873 >
INSERT < _123 mesur:hasNumericValue 
			(COUNT(?x) / COUNT(?y)) > .
\end{lstlisting}
\end{scriptsize}

As demonstrated, the presented metrics can be easily calculated using simple SPARQL queries. However, more complex metrics, such as those that are recursive in definition, can be computed using other semantic network algorithms. For example, the eigenvector-based Y-Factor \cite{journalstatus:bollen2006} can be computed in semantic networks using the grammar-based random walker framework presented in \cite{grammar:rodriguez2007}. The objective of the MESUR project is to understand the space of such metrics and their application to valuing artifacts in the scholarly community. Future work in this area will report the finding that are derived from such algorithms.

\section{Conclusion}

This article presented the MESUR ontology which has been engineered to provide an integrated model of bibliographic, citation, and usage aspects of the scholarly community. The ontology focuses only on that information for which large-scale real world data exists, supports usage research, and whose instantiation is scalable to an estimated 50 million articles and 1 billion usage events. A novel approach to data representation was defined that leverages both relational database and triple store technology. The MESUR project was started in October of 2006 and thus, is still in its early stages of development. While a trim ontology has been presented, the effects of this ontology on load and query times is still inconclusive. Future work will present benchmark results of the MESUR triple store.

\section{Acknowledgments}
This research is supported by a grant from the Andrew W. Mellon Foundation.

\balancecolumns
\end{document}